\documentclass[12pt,fleqn]{article}

\usepackage{epsfig}
\usepackage{amssymb}
\usepackage{amsthm}
\usepackage{amsmath}

\def\bib{\bibitem}

\newcommand{\rem}[1]{}
\newcommand{\sg}{\sigma_g}
\newcommand{\se}{\sigma_r}

\newcommand{\R}{{\mathbb{ R}}}

\newcommand{\UE}{U_{\rm eff}}

\newcommand{\fig}[1]{#1}

\begin{document}

\begin{titlepage}

\begin{centering}
{\bf\Large Generalizations of the St\"ormer Problem for \\ \vspace{1ex}
Dust Grain Orbits}
\vspace*{2cm}

H.R. Dullin  \\
Department of Mathematical Sciences, Loughborough University \\ 
Leicestershire, LE11 3TU, United Kingdom
\footnote{ email: h.r.dullin@lboro.ac.uk,
}
\vspace*{1cm}

M. Hor\'anyi and J. E. Howard \\
Laboratory for Atmospheric and Space Plasmas \\
University of Colorado, Boulder, CO 80309-0392

\end{centering}

\vspace*{3cm} \noindent {\bf Abstract:}
We consider the generalized St\"ormer Problem that includes the
electromagnetic and gravitational forces on a charged dust grain
near a planet. For dust grains a typical charge to mass ratio 
is such that neither force can be neglected. Including the
gravitational force gives rise to stable circular orbits 
that encircle that plane entirely above/below the equatorial 
plane. The effects of the different forces are discussed in
detail. A modified 3rd Kepler's law is found and analyzed 
for dust grains.

\vspace*{1ex} \noindent
PACS: 96.30.Wr, 45.50.Jf, 96.35.Kx 

\vspace*{1ex} \noindent
Keywords: Stormer Problem, Dust Grains, Halo Orbits, Stability 

\end{titlepage}

\section{Introduction}

One of the early milestones of space physics was St\"ormer's
theoretical analysis of charged particle
motion in a purely magnetic dipole field [1,2].  This seminal study
provided the basic physical framework that
led to the understanding of the  radiation belts surrounding the
Earth and other magnetized planets.
The radiation belts are now known to be composed of individual ions
and electrons whose motion is often
well described by magnetic forces alone.  These classical results are
also relevant to the dynamics of charged
dust grains in planetary magnetospheres.  However, the  much smaller
charge-to-mass ratios
produce a more complex dynamics, as planetary gravity and the
corotational electric field must also be taken
into account [3-6].

In a series of recent papers [7-10]  equilibrium and stability
conditions were derived for
charged dust grains orbiting about Saturn.  These orbits can be
highly non-Keplerian and
include both positively and negatively charged grains, in prograde or
retrograde orbits.
The first article was restricted to equatorial orbits, while the
second treated nonequatorial
``halo" orbits, i.e.\ orbits which do not cross the equatorial plane. 
Both assumed Keplerian gravity, an ideal aligned and
centered magnetic dipole
rotating with the planet, and concomitant corotational electric
field.  The third paper
dealt with the effects of planetary oblateness ($J_2$), magnetic
quadrupole field, and radiation pressure.
While the first two forces were found to have a negligible effect on
particle confinement, the
effects of radiation pressure could be large for distant orbits.
Interestingly, $J_2$ and
radiation pressure can act synergistically to select out one-micron
grains in the E-Ring [11].
The final paper in this series allowed the surface potential of a
grain, and hence its charge, to
adjust to local photoelectric and  magnetospheric charging currents.
It was concluded  that stable halo orbits were mostly likely to be
composed of rather small $(\approx 100nm)$ positively
charged grains in retrograde orbits.  A dust grain ``road map" was
drawn for the Cassini spacecraft now en route to Saturn,
showing where to expect dust grains of a given composition and radius.

This paper presents a more comprehensive treatment of dust grain
dynamics, but under the simplified assumptions of
Keplerian gravity, pure dipole magnetic field, and no radiation
pressure.  Some of the results were already presented in the letters [7,8],
here we fill in the details of the necessary calculation and also present
 new results.
Our goals are a mathematically rigorous
yet simplified derivation of equilibrium and stability  conditions
which highlights the relative importance of the
several different forces acting on an individual grain.

As is well known, there are no stable equilibrium circular orbits for 
the pure St\"ormer
problem of charged particle motion in a pure dipole magnetic field. 
It is the addition
of planetary gravity and spin that gives rise to stable families of 
equatorial and nonequatorial
orbits.
We begin with a general discussion of charged particle motion in 
axisymmetric geometry,
which is then specialized to the motion of charged grains in a 
planetary magnetosphere.
Equilibrium conditions are derived first for equatorial orbits, then 
for halo orbits.
Next we take up the issue of stability for each family of equilibrium orbits.
Results are presented for four distinct problems: the {\it Classical 
St\"ormer Problem} (CSP)
in which a charged particle moves in a pure dipole magnetic field, 
the {\it Rotational
St\"ormer Problem} (RSP), with the  electric field due to planetary 
rotation included,
the {\it Gravitational St\"ormer Problem} (GSP), with Keplerian 
gravity included but
not the corotational electric field, and the full system (RGSP) including both
fields.
For each case one must also consider each charge sign in prograde or 
retrograde orbits.
Our results may be summarized as follows:
\begin{description}
\item[CSP:] As is well known, no stable circular orbits, equatorial 
or nonequatorial exist.  However,
under adiabatic conditions important families of guiding center 
orbits confined to a potential trough
called the {\it Thalweg} exist. Such trajectories lie outside the 
scope of the present paper.

\item[RSP:]  Stable equatorial equilibria exist for both charge signs.
There are no halo orbits.

\item[GSP:] Stable equatorial equilibria exist for both charge signs.
Positive halos are retrograde and negative halos are prograde.
Both types are stable wherever they exist.

\item[RGSP:] Stable equatorial equilibria exist for both charge signs.
There is a range of positive charge-to-mass ratios without stable 
equatorial equilibria.
Negative halos are prograde, while positive halos can be pro- or retrograde.
For stability the frequency must be sufficiently different from twice the
rotation rate of the planet.
\end{description}
Therefore halo orbits do exist with and without the corotational 
electric field.
However, the corotational electric field is required in order to sustain stable
positive prograde halos. 

\section{Charged particle dynamics in axisymmetric geometry}

The equations of motion of a particle  of mass $m$ and charge
$q$ in $\R^3$,
${\bf r} = (x,y,z)^t$, are
\[
        m \ddot {\bf r} = \frac{q}{c}{\bf B} \times {\dot {\bf r}} -
\nabla U({\bf r})\,,
\]
where the potential $U(\bf r)$ generates the forces of gravity and perhaps
corotational electric field.
Denote by $R$ a rotation around the $z$ axis and assume that the
magnetic field ${\bf B}$ and the potential $U$
are symmetric with respect to this rotation:
\[
       {\bf B}( R {\bf r} ) = R {\bf B(r)}, \quad
       \nabla U(R {\bf r}) = R \nabla U({\bf r})\,.
\]
In particular this is true for the field ${\bf B = \nabla \times A}$
of a  centered magnetic dipole of strength ${\cal M}$ and dipole axis
the $z$ axis,
for which the  vector potential is, in the Coulomb gauge,
\[
      {\bf A(r)} = {\cal M}(y,-x,0)^t/r^3, \quad r^2 = x^2+y^2+z^2 \,.
\]

The equations of motion can be transformed to a rotating coordinate system
using a rotation matrix $R$ corresponding to the angular velocity 
${\bf \Omega} = (0,0,\Omega)^t$ which rotates around the $z$-axis
with with angular speed $\Omega$.
For given angular velocity $\bf \Omega$ the $z$ axis is
chosen in the same direction, so that $\Omega$ is positive. Note that the magnetic
moment $\cal M$ can
be positive or negative.
Direct differentiation then gives
\[
     {\bf r} = R {\bf q}, \quad
     \dot {\bf r}  = R({\bf \dot q + \Omega \times q}), \quad
     \ddot {\bf r} = R({\bf \ddot q} + 2 {\bf \Omega \times \dot q +
\Omega\times(\Omega\times q)}) \,.
\]
so that
\[
     m \ddot {\bf q} = (\frac q c {\bf B} - 2 m {\bf \Omega}) \times
\dot {\bf q} - m {\bf \Omega \times ( \Omega\times q) }
       + \frac q c {\bf B \times (\Omega\times q)} - \nabla U({\bf q}) \,,
\]
where  $-2m{\bf \Omega} \times \dot {\bf q}$ is the Coriolis
force and
$-m{\bf \Omega \times (\Omega\times q)}$ is the centrifugal force.
Following the standard argument  the term $\frac q c {\bf B
\times (\Omega \times q)}$ is not present
in a frame rotating with the planet, because there is no additional
electric field.
Therefore there must be the same term in $\nabla U$ in order to cancel it.
In particular we model the situation inside the corotating plasma,
for which this assumptions is reasonable.
Transforming back to the rest frame this addition to the potential
gives the corotational electric field observed in the rest frame.
The corresponding potential is obtained from
\[
       {\bf E} = \frac q c {\bf B \times (\Omega \times r)} = \gamma
\Omega \nabla \Psi,
\quad \Psi = \frac{x^2+y^2}{r^3} \,,
\]
where $\gamma = q{\cal M}/c$.
The electric field ${\bf E}$
is unipolar, i.e. its curl is zero, and therefore it
is not induced by a changing magnetic field.
Moreover, it is perpendicular to the magnetic field, ${\bf E \cdot B} = 0$.
The divergence of this field is not zero;
instead we find $\nabla \cdot {\bf E} = 2\gamma \Omega (2z^2-x^2-y^2)/r^5$,
so that there is a space charge distribution originating from
the rotation of the plasma.

In an inertial frame the potential now reads
\[
      U({\bf r}) = -\sg\frac{\mu m }{r} + \se\gamma \Omega \Psi \,,
\]
where the parameters $\sg$ and $\se$ serve as markers of the
gravitational and electric
forces in order to track the origin of the various terms after
scaling away excess parameters.
Usually we consider $\se = \sg = 1$; the classical St{\"o}rmer
problem [1] 
has $\se = \sg = 0$. The case $\se = 0$ and $\sg=1$ has also been studied
by St{\"o}rmer [1,2]; 
we shall refer to it as the ``Gravitational St{\"o}rmer Problem." While it is
simpler than our case,
   it will turn out that the most important physical effects can
already be seen in this subcase.
Another interesting special case is $\sg = 0$ and $\se = 1$. This
takes into account the
effect of the electric field in the rotating plasma, but the particle
is still massless.
We call this case the ``Rotational St{\"o}rmer Problem." It will turn
out that in this case halo orbits do not exist.

The Hamiltonian of the above equations of motion is
\begin{equation} \label{eqn:Hxyz}
       H = \frac{1}{2m}\left( {\bf p} - \frac{q}{c} {\bf A(r)}\right)^2
+ U({\bf r})\,.
\end{equation}
Owing to the symmetry of the problem the $z$ axis is an invariant
set. Because it
is singular in coordinates adapted to the symmetry of the problem
it is best analyzed in cartesian coordinates.
The magnetic field is parallel to this axis, so that for motion on this
axis there
is no Lorentz force. For initial conditions in the set  $x=y=p_x=p_y=0$
the equations of motion in cartesian coordinates show that the derivatives of
$x,y,p_x,p_y$ are zero; hence it is an invariant set, on which the Hamiltonian
is purely gravitational, $H_z = p_z^2/2m - \mu m / z$. Depending on
the initial conditions in the invariant set a particle either collides with
the planet or escapes to infinity; there are no stationary points on this axis.

In cylindrical coordinates
$(\rho=\sqrt{x^2+y^2},\phi=\arctan(y,x),z)$ $H$ becomes
\begin{equation} \label{eq:Hcyl}
       H = \frac{1}{2m}\left( p_\rho^2 + p_z^2 + \frac{1}{\rho^2}(p_\phi
- \gamma \Psi)^2 \right)
            -\sg\frac{\mu m }{r} + \se \gamma \Omega \Psi \,,
\end{equation}
with the dipole stream function $\Psi = \rho^2/r^3$.
For systems with $S^1$ symmetry a stream function $\Psi$
independent of $\phi$
can always be introduced.
The corresponding vector potential is $(\sin\phi,-\cos\phi,0)^t\Psi/\rho$,
the magnetic field is $(x \Psi_z, y \Psi_z, -\rho \Psi_\rho)^t/\rho^2$,
and the electric field is $\Omega(x\Psi_\rho, y\Psi_\rho,\rho \Psi_z)^t/\rho$.

\rem{
\footnote{
Instead of formulating the Hamiltonian in the rest frame with the
additional ${\bf E}$ field we can also find a Hamiltonian in
the rotating frame with no ${\bf E}$ field but modified ${\bf A}$
and additional centrifugal term.
The centrifugal term can also be absorbed into the potential using
\[
       {\bf \Omega \times (\Omega \times r)} = \Omega^2 \nabla (x^2 + y^2)/2 \,.
\]
The vector potential of the constant coriolis field $-2 m {\bf
\Omega}$ is given by
${\bf A}_c = m{\bf \Omega} \times {\bf r} = m \Omega (y,-x,0)^t$.
Therefore the Hamiltonian that generates the equations of motion in
the rotating
frame reads
\[
       H_r = \frac{1}{2m}\left( {\bf p} - \frac{q}{c} {\bf
A}_r\right)^2 + U_r \,,
\]
where ${\bf A}_r = {\bf A + A}_c$ and $U_r = -\mu m/r + \Omega^2 \rho^2/2$;
in cylindrical coordinates:
\[
      H_r = \frac{1}{2m}\left( p_\rho^2 + p_z^2 + \frac{1}{\rho^2}
(p_\phi - \gamma\Psi + m\Omega \rho^2)^2 \right) + U_r\,.
\]
We will not use this Hamiltonian.
}
}

We may distinguish three types of constants in the problem:
\begin{itemize}
       \item Parameters describing the planet's mass $\mu=GM$ and spin
rate $\Omega$.
       They are the most fixed parameters.
       \item Parameters describing the dust particle's mass $m$ and
charge, measured by
       $\gamma = q {\cal M}/c$.
       \item The angular momentum $p_{\phi}$ and total energy $h=H$ are
       constants of the motion determined by the initial
       conditions.  Fixing both $h$ and $p_\phi$ defines a region of possible
       motion in configuration space.
\end{itemize}

We now introduce a convenient scaling to reduce the number of parameters.
Time is measured by the inverse frequency of
the planetary spin rate $\Omega$.
Distances are measured in terms of the radius of
the Keplerian synchronous orbit
\[
     R = (\mu/\Omega^2)^{1/3} \,
\]
while mass is measured in units of the particle mass $m$. The scaled
Hamiltonian is then
\begin{equation} \label{eqn:Hscal}
     \hat H = \frac12\left( \hat p_\rho^2 + \hat p_z^2 +
       \left( \frac{\hat p}{\hat\rho} - \delta \frac{\hat\rho}{\hat r^3}
\right)^2 \right)
       - \frac{\sg}{\hat r} + \se \delta \frac{\hat\rho^2}{\hat r^3} \,,
\end{equation}
where the variables with hat are measured in the new scale.
   From now on we drop all hats.
The essential dimensionless parameters are
\[
	p = p_\phi \frac{R\delta}{\gamma} \qquad \text{and} \qquad
    \delta = \frac{\Omega\gamma}{m\mu} = \frac{q}{m} \frac{\cal M}{c}
\frac{\Omega}{G M}= {{\omega_c \Omega}\over{\omega_k^2}}
\]
where $\omega_c = qB_0/mc$ is the cyclotron frequency, $B_0$ is the planetary magnetic field on the equator,  
$\omega_k = \sqrt{GM/R_s^3}$ is the Kepler frequency, with $R_s$ the planetary radius, 
and the parameter $p$ is just the angular momentum $p_\phi$ measured in 
the new units.
The single parameter for the dust grain is $\delta$,
which is essentially the charge-to-mass ratio.
Recall that the $z$ axis is oriented so that $\Omega > 0$.
In the following we will loosely talk about positive/negative charge
when we mean
positive/negative $\delta$. This correspondence is correct if the
magnetic dipole
moment $\cal M$ is positive, i.e. the spin and the field are aligned. This is
true for Saturn, the main application that we have in mind.
Our results are valid in both cases.

\section{Equilibria}
\subsection{Equatorial Orbits}

Here we shall find it advantageous to work in
spherical coordinates
$(r=\sqrt{\rho^2+z^2},\theta=\arccos(z/r),\phi)$, rather than the
cylindrical coordinates of Ref [7,8]. 
The Hamiltonian becomes
\begin{equation} \label{eq:Hspher}
			H =
\frac{1}{2}\left(p_r^2+\frac{p_\theta^2}{r^2}\right) + U_{\rm eff} \,,
\end{equation}
where the effective potential $U_{\rm eff}$ is the part of the Hamiltonian
independent of the non-constant momenta $p_r$ and $p_\theta$:
\begin{equation} \label{eq:Ueff}
    U_{\rm eff}(r,\theta,p) = \frac{(p r-\delta\sin^{2}\theta)^{2}}
		{2 r^{4}\sin^{2}\theta}
           - \frac{\sg - \se\delta\sin^{2}\theta}{ r} \,.
\end{equation}

The equations of motion are then
\begin{eqnarray}
&&  \dot r = p_r, \quad \dot \theta = p_\theta, \quad \dot \phi =
\partial_p U_{\rm eff}, \\
&&  \dot p_r = -\partial_r U_{\rm eff}, \quad
       \dot p_\theta = -\partial_\theta U_{\rm eff}, \quad
       \dot p_\phi = 0 \,.
\end{eqnarray}
In order to facilitate the calculation of the partial derivatives of $\UE$
we introduce the frequency
\begin{equation} \label{eq:omega}
	\omega(r,\theta) = \dot\phi = \partial_p U_{\rm eff}  =
\frac{p}{r^2\sin^2\theta} -
	  	\frac{\delta}{ r^3} \,.
\end{equation}
In analyzing  circular equilibrium orbits it is preferable to employ
$\omega$ rather
than $p_{\phi}$ as parameter, as it is the sign of $\omega$ that determines
whether the orbit is rotating in the same direction as the planet
(prograde) or opposite to it (retrograde).
Recall that in the scaled variables frequencies are measured in
terms of $\Omega$; hence $\omega=1$  means synchronous motion.
   From now on we will eliminate $p$ in favour of
$\omega = \omega(r,\theta)$ in the potential to get
\begin{equation}
	\UE = \frac12 \omega^2 r^2 \sin^2\theta - \frac{\sg}{r}
		+\frac{\se\delta\sin^2\theta}{r} \,.
\end{equation}
It is important to notice that when calculating derivatives of $\UE$
with respect to $r$ and $\theta$ we have to treat $\omega$ as a
function of $r$ and $\theta$.
The derivatives of $\omega$ (re-expressed in terms of $\omega$) are
\begin{eqnarray*}
     \partial_r \omega &=& -2\frac{\omega}{r} + \frac{\delta}{r^4} \\
     \partial_\theta\omega &=& -2 \cot\theta
		\left( \omega + \frac{\delta}{r^3}\right) \, ,
\end{eqnarray*}
so that
\begin{eqnarray}
	\label{eq:drU}
     \partial_r \UE &=& -\omega^2r\sin^2\theta + \frac{1}{r^2}
	(\delta(\omega-\se)\sin^2\theta + \sg) \\
	\label{eq:dtU}
     \partial_\theta \UE &=& -\frac{1}{r} \cos\theta\sin\theta
	(\omega^2r^3+2\omega\delta-2\se\delta) \,.
\end{eqnarray}

If all partial derivatives with respect to $r$, $\theta$, and $p$ are zero
there is no motion at all. The latter is just $\omega$, which we set to zero.
Then (\ref{eq:dtU}) requires $\theta = \pi/2$. The other possibility
$\theta = 0,\pi$ is the coordinate singularity, and it has already been
treated using $H_z$. The uncharged case $\delta = 0$ is not
of interest here.  For $\theta = \pi/2$ and $\omega=0$ (\ref{eq:drU}) reduces
to $\sg = \se\delta$. Therefore particles at rest can occur anywhere in the
equatorial plane but only when $\delta = \sg/\se$. Therefore $\delta = 1$
can be considered as the case where electrical and gravitational forces are
balanced.
In the classical St{\"o}rmer problem $\se = \sg = 0$ both equations
are automatically
satisfied once $\omega=0$. In this case we can place a particle at
rest anywhere
in space; since there is no potential there are no forces if there is
no motion.
In any case it is true that for these trivial solutions at rest the
angular momentum $p$ is nonzero; from $\omega = 0$ and (\ref{eq:omega}) we
find $p = \delta/r$.  The energy is zero for all these equilibrium points.

The system has a discrete symmetry: the equations of motion are invariant
under the map $(\theta,p_\theta) \to (\pi - \theta, -p_\theta)$.
The set that is invariant under this map is the equatorial plane with
no transverse momentum, $(\theta,p_\theta) = (\pi/2, 0)$. As always this
is also an invariant set for the dynamics.
The physical reason for this is that ${\bf B(r)}$ is
parallel to the $z$ axis if ${\bf r} = (x,y,0)$, so that for motion
within the equatorial plane there is
a Lorentz force, however, with direction in the plane.
Moreover, ${\bf E(r)}$ is perpendicular to ${\bf B(r)}$, and
therefore ${\bf E(r)}$ lies in
the equatorial plane  if ${\bf r} = (x,y,0)$.
The Hamiltonian restricted to the equatorial plane reads
\[
      H_{xy} = \frac{1}{2} \left( p_r^2 +
\frac{1}{r^2}\left(p-\frac{\delta}{r}\right)^2\right) -
         \frac{\sg - \se\delta}{r} \,.
\]
This is an integrable system with one effective degree of freedom
that can be solved in terms of elliptic functions.
The effective potential in the equatorial plane is
\[
       U_{xy}(r) = \frac12 \left(
\frac{p}{r}-\frac{\delta}{r^2}\right)^2   - \frac{ \sg -
\se\delta}{r}  =
                \frac12 \omega^2 r^2 - \frac{ \sg - \se\delta}{r} \,.
\]
The minima $r_e$ of $U_{xy}(r)$ correspond to circular orbits in the
equatorial plane
because the right hand sides of all the equations of motion except
$\dot \phi = \omega$
are zero. The calculation of critical points of $U_{xy}(r)$ leads to
the solution of a cubic polynomial in $r$ given by
$r^2 \partial_r U_{xy}$. If instead we eliminate $p$ in favour of $\omega$
we obtain a much simpler polynomial
\begin{equation} \label{eqn:E}
        P(r,\omega)  = \omega^2 r^3 - \omega\delta  + \se\delta - \sg \,.
\end{equation}

\begin{figure}
\fig{   \centerline{\epsfxsize=2in\epsfbox{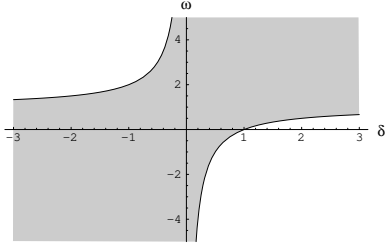}}
}		\caption{\small{Existence of equatorial orbits}}
     \label{fig:Eexist}
\end{figure}

Solving $P=0$ for $r$ yields a generalization of Kepler's 3rd law
for equatorial orbits:
\begin{equation}
	\label{eq:re}
	r_e(\omega)^3 = \frac{\sg + \delta(\omega-\se)}{\omega^2} \,.
\end{equation}
which for $\delta \to 0$ reduces to the ordinary Kepler's law.
The corresponding angular momentum $p$ can be calculated from
(\ref{eq:omega}) and is
$p_e = \omega r_e^2 - \delta/r_e$.
The radius $r_e$ is positive if
\begin{eqnarray} \label{eqn:eqex1}
	\omega \le \se - \frac{\sg}{\delta} &\text{ and }& \delta \le
0, \qquad \text{or} \\ \label{eqn:eqex2}
	\omega \ge \se - \frac{\sg}{\delta} &\text{ and }& \delta \ge 0 \,.
\end{eqnarray}
For negative $\delta$ there are always pro- and retrograde orbits.
For positive $\delta$ this is only true for $\delta < \sg/\se$, while
for $\delta > \sg/\se$ all orbits are prograde.
In Fig.~\ref{fig:Eexist} the possible combinations of $\omega$ and $\delta$
for which circular equatorial orbits exist are shaded grey.
The horizontal asymptotes have $\omega = \se$.
The hyperboloidal boundaries and the limit $\omega \to \infty$
correspond to zero radius.
In the St{\"o}rmer case $\se = \sg = 0$ all equatorial circular orbits
of negatively charged particles ($\delta < 0$) are retrograde ($\omega < 0$),
while positively charged particles have prograde orbits. The gravitational
and/or electric field perturbations create small regions of the opposite
behaviour for some $\delta$. In addition some motions for large positive
$\delta$ and small positive $\omega$ are made impossible by switching on the
additional fields.

\begin{figure}
\fig{   \centerline{\epsfig{file=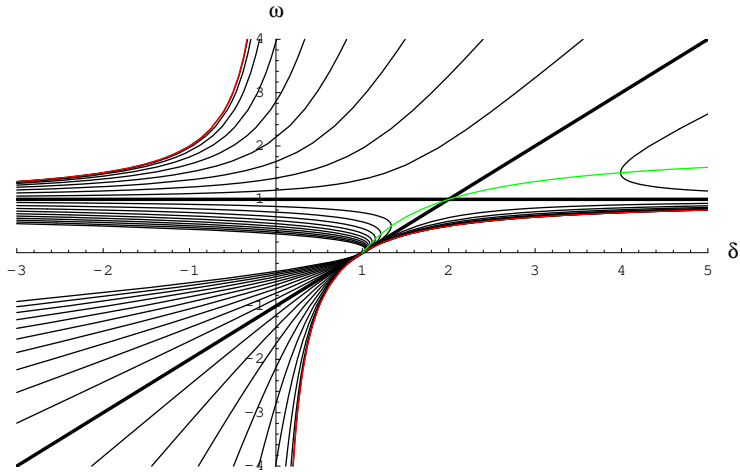,width=4in} }}
		\caption{\small{Curves of constant radius $r=0.1 i$,
$i=1,\dots,20$ for equatorial orbits.
      The thick lines have $r=1$.}}
     \label{fig:Ercon}
\end{figure}

\begin{figure}
\fig{   \centerline{ \epsfig{file=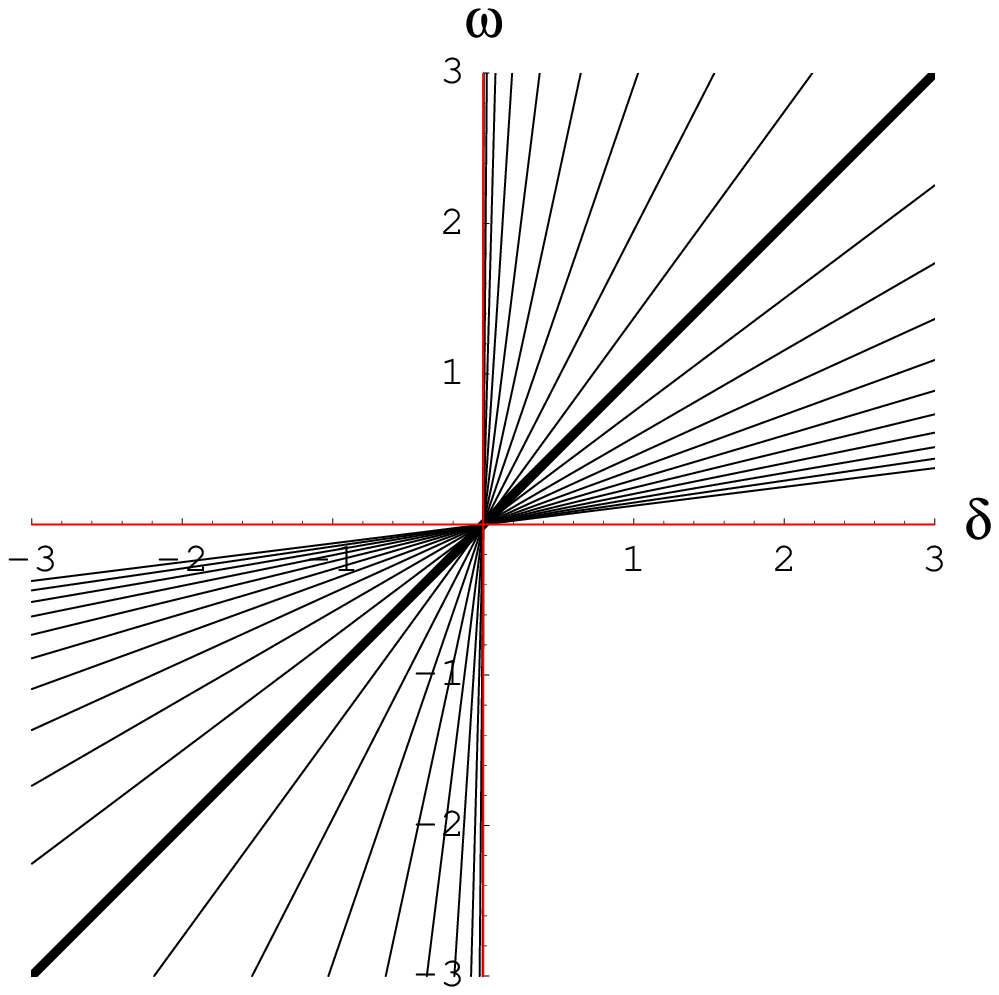,width=2in} $\quad$
\epsfig{file=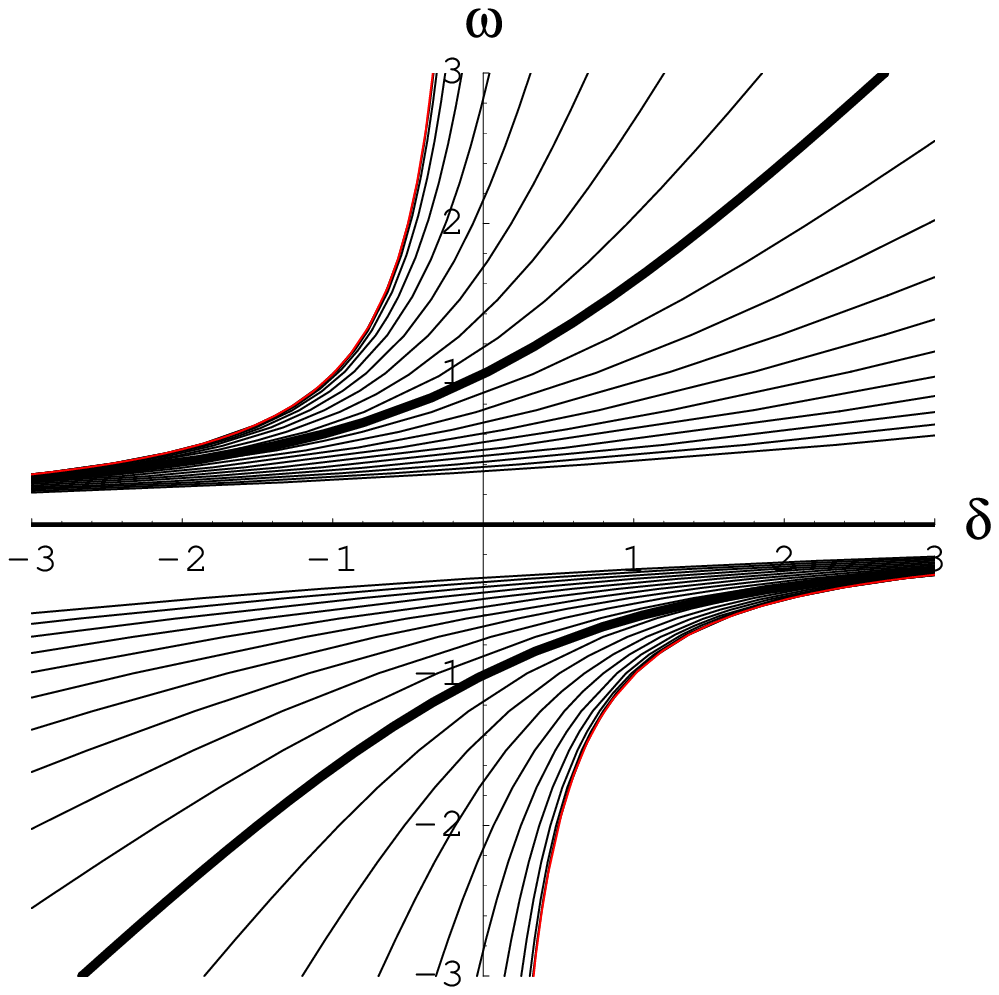,width=2in} $\quad$
\epsfig{file=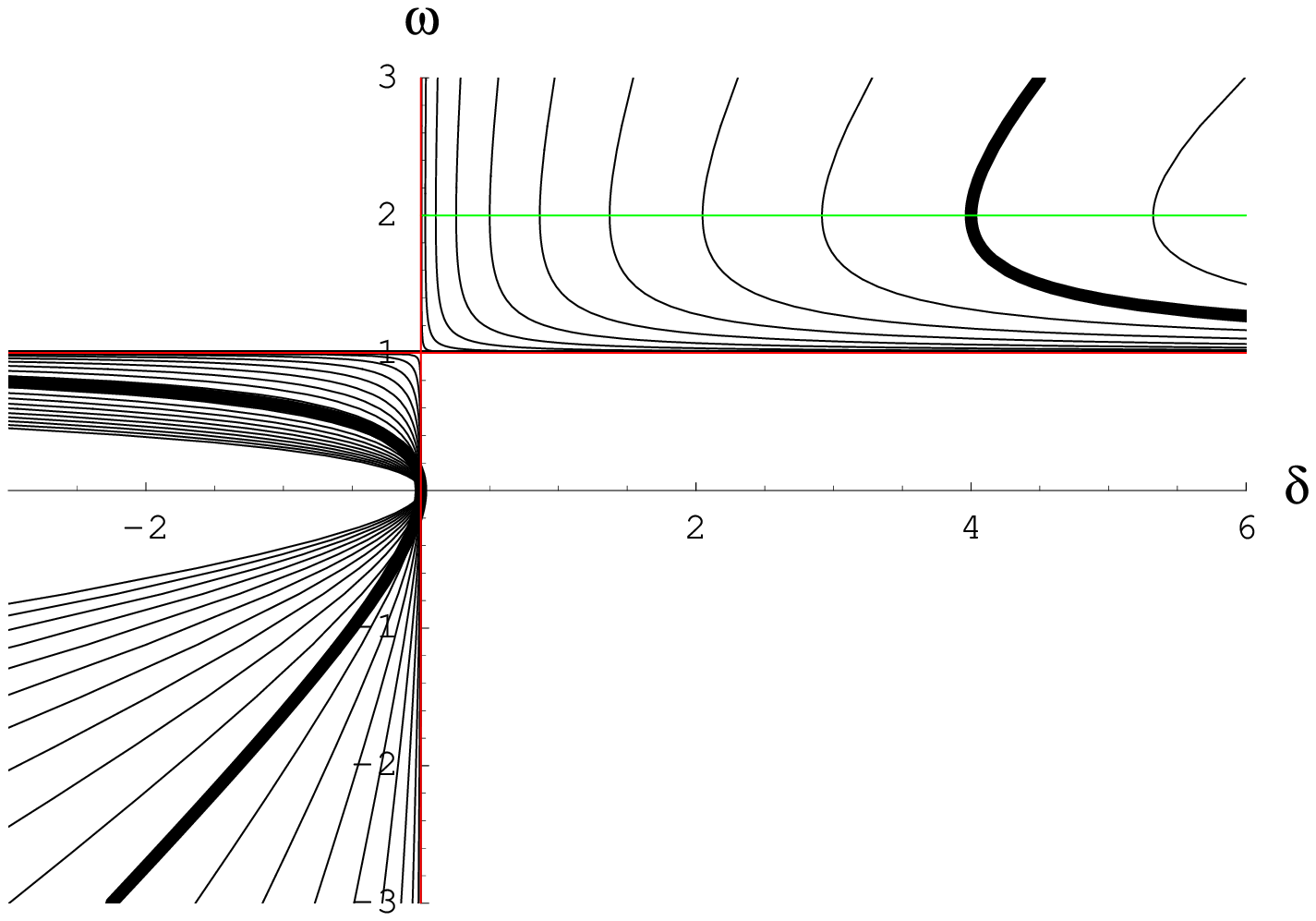,width=3in} }
}
		\caption{\small{Curves of constant radius for
equatorial orbits of the
a) classical ($\se=\sg=0$), b) gravitational ($\se=0,\sg=1$), and c)
rotational ($\se=1,\sg=0$) St{\"o}rmer problem.
Thick lines have $r=1$.}}
     \label{fig:ESrcon}
\end{figure}

The equation for equatorial orbits, $P=0$, can be solved for $\delta$
in order to give $\delta$ as a function
of $\omega$ for given $r$:
\[
     \delta = \frac{r^3\omega^2 - \sg}{\omega-\se} \,.
\]
These curves are shown in Fig.~\ref{fig:Ercon} and \ref{fig:ESrcon}.
Note the two straight lines with $r=1$,
which in our scaling is the radius of the synchronous orbit in the
Kepler problem.
The horizontal one corresponds to the synchronous orbits ($\omega=1$) which
exist for any $\delta$. Hence the synchronous Kepler orbit is not
affected by the
addition of both fields. This is not true for the three St{\"o}rmer
cases, see below.
Another prominent feature is the point $(\delta,\omega)=(1,0)$, which
is intersected
by hyperbolas with all $r$. It corresponds to the equilibrium points 
discussed above.

Let us briefly discuss the corresponding diagrams for the three
St{\"o}rmer cases
shown in Fig.~\ref{fig:ESrcon}.
In the classical case $\omega$ and $\delta$ are proportional. Small slope means
large $r$. The relation breaks down for the line $\omega=0$ for which all radii
are possible. There are only two other types of equatorial orbits: 
positively charged
prograde and negatively charged retrograde.
In the gravitational St{\"o}rmer case the relation between $\omega$
and $\delta$
is quite similar, but now there are also small regions of the two types
of motions: positively charged retrograde and negatively charged prograde.
In the rotating St{\"o}rmer case the positive retrograde orbits have
disappeared again. The main new feature is the appearance of orbits of small
negative charge with small $\omega$ and small radius.

\subsection{Halo Orbits}

Our goal is the calculation of periodic orbits that encircle the planet
in a plane parallel to the equatorial plane but entirely above/below it.
Circular orbits correspond to critical points of $U_{\rm eff}$,
i.e. points $(r_{0},\theta_{0})$ at which both derivatives of $U_{\rm
eff}$ vanish.
This is so because at $(r,\theta,p_r,p_\theta) = (r_0,\theta_0,0,0)$
the right hand sides of Hamilton's equations are zero, except for $\dot\phi$.
They are given by the minima of $U_{\rm eff}$.
Their stability will be calculated in the next section.

Circular orbits are given by the solution of $\partial_r U = 0$
and $\partial_\theta U = 0$ for arbitrary $\omega$, see
(\ref{eq:drU},\ref{eq:dtU}).
The second equation has the solution
$\theta = \pi/2$, which gives the equatorial orbits we already analyzed.
Also the solutions with $\theta = 0,\pi$ have already been described using the
reduced Hamiltonian $H_z$.
The remaining solutions are given by $Q=0$ with
\begin{equation} \label{eq:H}
        Q(r,\omega) = \omega^2r^3 + 2\omega\delta - 2\se\delta \,,
\end{equation}
which describe the nonequatorial (or halo) orbits.
The equation $Q=0$ can be solved for $r^3$, which
can then be eliminated from (\ref{eq:dtU}) resulting in
an angular equation $A=0$ with
\begin{equation} \label{eq:A}
        A(\theta,\omega) = \sg + 3\delta(\omega-\se)\sin^2\theta \,.
\end{equation}
The functions $Q$ and $A$ completely describe the
halo orbits.
In particular these equations can easily be solved
for $r$ and $\theta$, so that all circular orbits
are obtained in parametric form, with $\omega$ as a parameter.
Explicitly we find
\begin{eqnarray}
\label{eq:r3is}
    	r_h(\omega)^3 &=& 2\delta \frac{\se-\omega}{\omega^2} \\
\label{eq:s2is}
    	\sin^2\theta_h(\omega) &=& \frac{\sg}{3\delta(\se-\omega)} \, .
\end{eqnarray}
grains [4]. 
The second equation clearly shows that without the gravitational
forces ($\sg = 0$) there are no halo orbits.
In particular in the classical St{\"o}rmer problem there are
no halo orbits (except for the trivial equilibrium points discussed above).
Adding the electric field alone there still are no halo orbits.
In both equations
it is obvious that the electric field merely  shifts
the frequency $\omega$ of circular halo orbits. We conclude that the
electric field is not essential for the existence of halo orbits.

\begin{figure}
\fig{   \centerline{\epsfxsize=2in\epsfbox{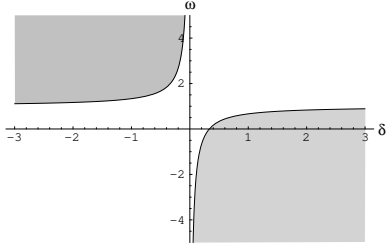}}}
		\caption{\small{Existence of halo orbits}}
     \label{fig:Hexist}
\end{figure}

For halo orbits the essential condition for their existence is that
$\sin^2\theta \le 1$,
which implies
\begin{eqnarray}
	\omega \ge \se - \frac{\sg}{3\delta} &\text{ and }& \delta
\le 0, \qquad \text{or}\\
	\omega \le \se - \frac{\sg}{3\delta} &\text{ and }& \delta \ge 0 \,.
\end{eqnarray}
These conditions automatically imply that the corresponding radius
$r_h$ is positive. Note that the ordering of the $\omega$ inequalities
is reversed compared with the equatorial case 
(\ref{eqn:eqex1},\ref{eqn:eqex2}).
Hence for negative
charge  only  prograde orbits exist while for small positive charge
 only  retrograde orbits exist.
Only if $\se \not = 0$ and for $\delta>\sg/3 \se$ can both types of
orbits exist.
The electric field does make a difference for the existence of synchronous
halo orbits: If $\se=1$ then $\omega = 1$ is impossible for finite $\delta$.
If $\se = 0$  synchronous halo orbits exist for $\delta < -\sg/3$.
Note that $\omega=\se$ (in particular a synchronous orbit if $\se=1$)
is impossible for finite $\delta$.
Without the electric field there do exist synchronous halo orbits
with negative charge.
In order for $(q/c) {\bf B} \times \dot {\bf r}$ to balance gravitation for
positive charge we have to reverse $\dot {\bf r}$, hence $\omega = -1$.

\begin{figure}
\fig{   \centerline{\epsfig{file=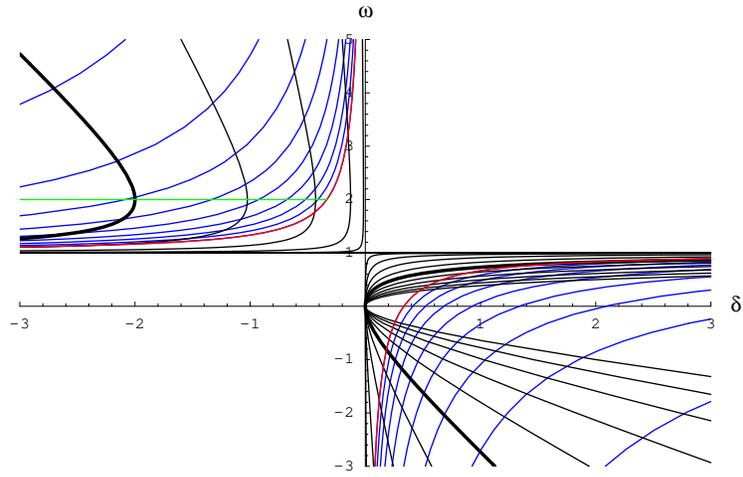,width=4in} }}
		\caption{\small{Curves of constant $r$ and $\theta$
for halo orbits.
Thick lines have $r=1$. $r = 0.2 i$,  $\sin\theta = 0.1 i$, $i=1,\dots,10$}}
     \label{fig:Hrcon}
\end{figure}

\begin{figure}
\fig{   \centerline{\epsfig{file=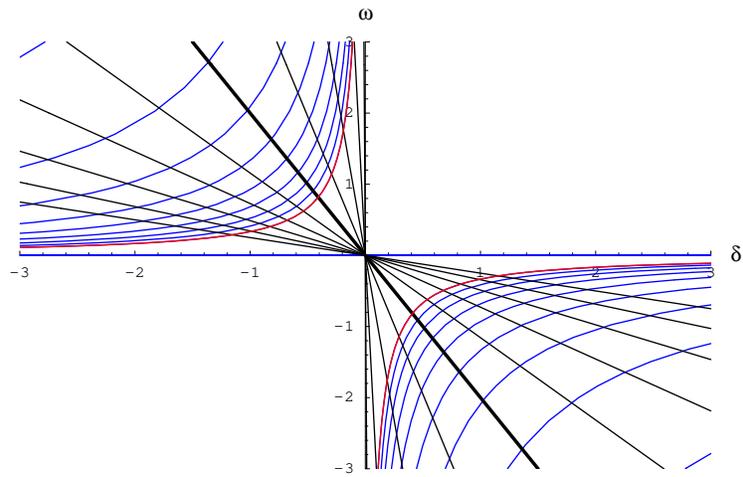,width=4in} }}
		\caption{\small{Curves of constant $r$ and $\theta$
for halo orbits in
the gravitational St{\"o}rmer problem. Thick lines have $r=1$.
$r = 0.2 i$,  $\sin\theta = 0.1 i$, $i=1,\dots,10$}}
     \label{fig:HSrcon}
\end{figure}

To get an overview of all possible halo orbits we plot curves of
constant $r$ and $\theta$ in the $\delta$-$\omega$ plane.
The family of  curves of constant radius is given by
\[
      \delta = \frac{r^3\omega^2}{2(\se - \omega)}.
\]
The family of curves of constant azimuth $\theta$ is given by
\[
      \delta = \frac{\sg}{3(\se-\omega)\sin^2\theta}.
\]
Both families are shown in Fig.~\ref{fig:Hrcon} for $\se=\sg=1$
and in Fig.~\ref{fig:HSrcon} for the St{\"o}rmer problem with
gravitation.

The regions of existence in the $\delta$-$\omega$-plane for
equatorial and halo orbits only overlap
in a small region bounded by hyperbolae. They are in a sense
connected at the hyperboloidal boundary of
the halo orbits, because in the next section we will find that this
line marks a pitchfork bifurcation
of an equatorial orbit changing its stability and creating halo orbits.

\section{Stability}
In Ref [7-8] 
explicit stability boundaries for both equatorial and halo orbits
were calculated.
Here we obtain these boundaries  more directly using the fact that
all circular orbits may be  parameterized by $\omega$.
A circular orbit is stable if it corresponds to a local minimum of $U$,
for which we need the second derivatives of $U$.
Using the chain rule gives
\begin{eqnarray*}
     \partial_r^2 U &=&
\left (3\,{\omega}^{2}-2\delta{\frac {3\,\omega-\se}{
{r}^{3}}}+{\frac {{\delta}^{2}}{{r}^{6}}}\right )
	\sin^2\theta - \frac{2\sg}{r^3} \\
     \partial_\theta^2 U &=&
\left (2\,{r}^{2}{\omega}^{2}+4\delta{\frac {\omega+\se}{r}}
         +4\,{\frac {{\delta}^{2}}{{r}^{4}}}\right )\cos^2\theta
+{r}^{2}{\omega}^{2}
      +2\delta {\frac { \omega-\se }{r}} \\
     \partial_r \partial_\theta U & = &
2\,\left ( r{\omega}^{2}
     +  \delta {\frac {2\omega-\se}{{r}^{2}}}
-{\frac {{\delta}^{2}}{{r}^{5}}}
\right )\cos\theta\sin\theta.
\end{eqnarray*}

\subsection{Equatorial Orbits}
\begin{figure}
\fig{   \centerline{\epsfig{file=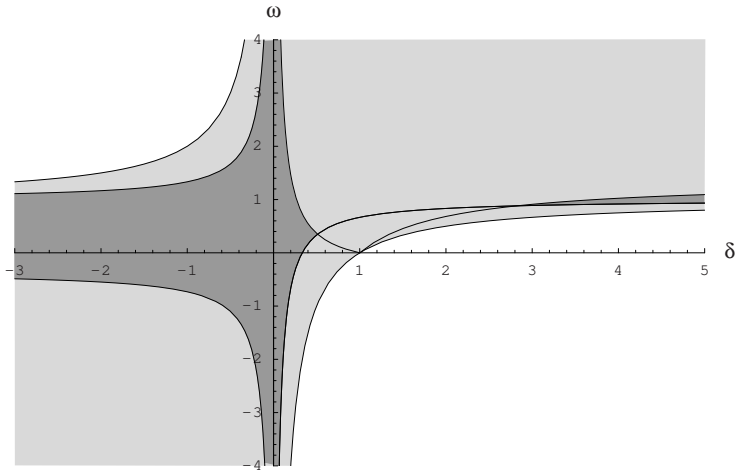,width=4in}}}
		\caption{\small{Regions of stability for equatorial
orbits are dark. Overlay Fig.~\ref{fig:Ercon}}}
     \label{fig:Estable}
\end{figure}

For equatorial orbits we insert $\theta = \pi/2$ and $r = r_e(\omega)$
as given above. The mixed derivative vanishes and the other two are
\begin{eqnarray*}
     \partial_r^2 U |_e &=& -\frac12
	\left(\frac{\omega}{\sg+\delta(\omega-\se)}\right)^2
	((2\omega\delta - \se\delta+\sg)^2 - 3(\se\delta - \sg)^2) \\
     \partial_\theta^2 U |_e &=& \sg + 3\delta(\omega-\se) \,.
\end{eqnarray*}
The radial derivative diverges for those $\omega$ that correspond
to $r=0$. It vanishes for $\omega=0$; however, the corresponding
radius $r_e$ is not finite (except for $\delta=\sg/\se$, the case of
equilibrium points).
There are two nontrivial factors that correspond to
tangent bifurcations of equatorial orbits with
\[
\omega_e^\pm = \frac{1\pm\sqrt{3}}{2}\left(\se - \frac{\sg}{\delta}
\right) \, .
\]
The vanishing of the
second $\theta$ derivative indicates the loss of transverse stability.
Because of reflection symmetry $\theta \to -\theta$
this results in a pitchfork bifurcation with
\[
     \omega_{PF} = \se-\frac{\sg}{3\delta}. \qquad
\]
All three curves are
hyperbolas in the space of $(\delta,\omega)$
shown in Fig.~\ref{fig:Estable}. The intersection of the curves
$\omega_{PF}$ and $\omega_e^\pm$ occurs at
\[
      \delta_e^\pm = \frac\sg\se \frac{5\pm 2\sqrt{3}}{3} \,.
\]


Using the above formulas for the second derivatives it is easy
to check that stability only holds in the following $\omega$ ranges
(compare~Fig.~\ref{fig:Estable}).
\begin{eqnarray*}
	\delta < 0 & : & \omega_e^- < \omega < \omega_{PF} \\
	0 < \delta < \delta_e^- & : & \omega_{PF} < \omega < \omega_e^- \\
	\delta > \delta_e^+ & : & \omega_{PF} < \omega < \omega_e^+ \,.
\end{eqnarray*}
For the first two ranges $\omega=0$ is (partially) included, which means
that the corresponding family of orbits exists for arbitrary
large radius. The radius $r_e$ as given by (\ref{eq:re})
as a function of $\omega$ is a monotone
function for most of these orbits except for unstable orbits
with $\delta > 1$. This gives the additional curve in
the diagram. The fact that the radius is not monotonic can already
be seen in Fig.~\ref{fig:Ercon}, where the turning points are marked
by a gray line.

\begin{figure}
\fig{   \centerline{\hfill \epsfig{file=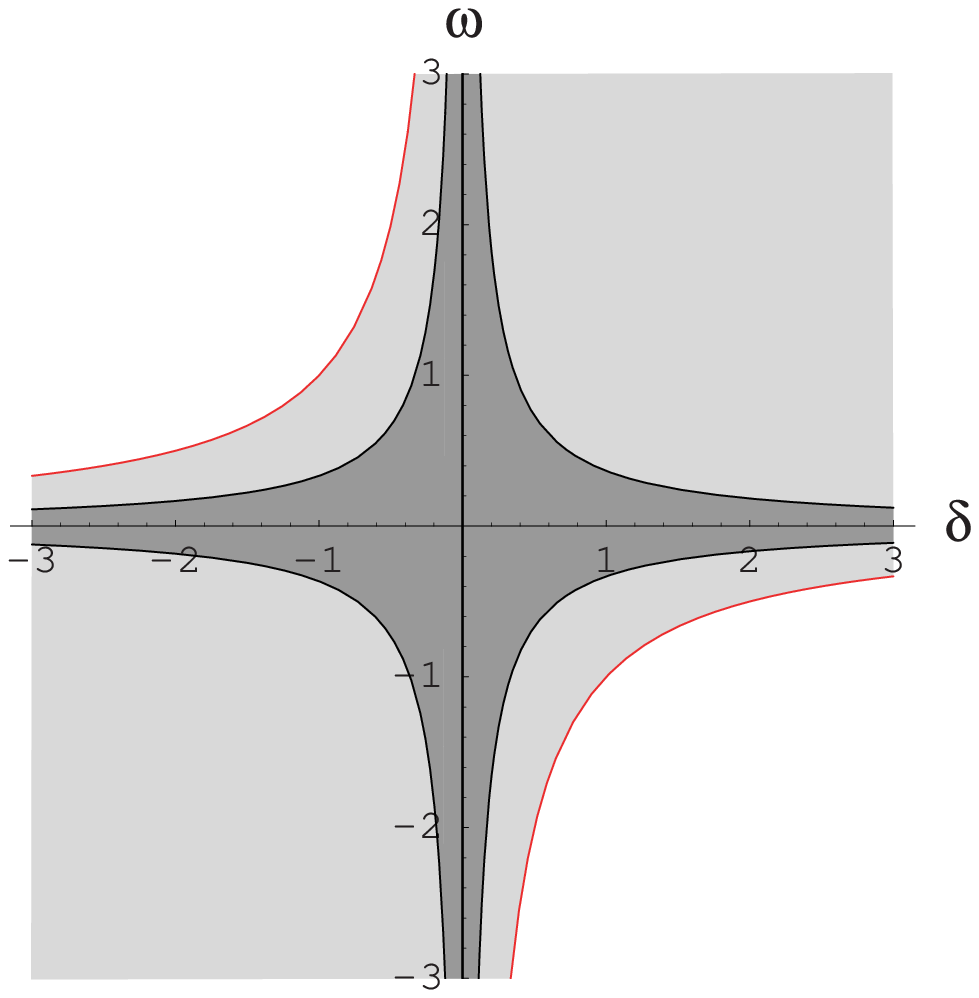,width=2in} $\quad$
  \epsfig{file=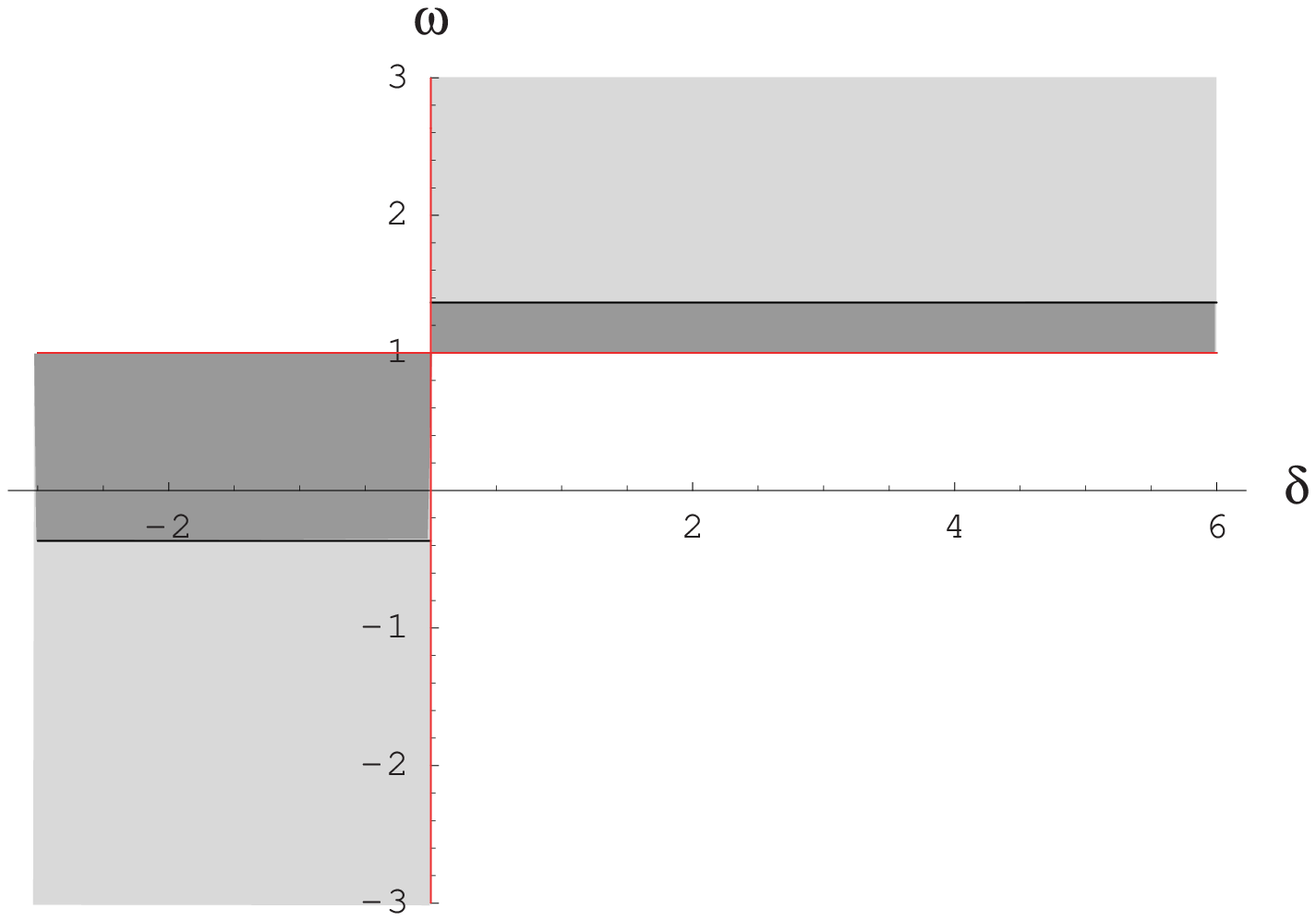,width=3in} \hfill}}
		\caption{\small{Regions of stability for equatorial
orbits of the
a) gravitational, and b) rotating St{\"o}rmer problem are dark.
Overlay Fig.~\ref{fig:ESrcon}}}
     \label{fig:Sstable}
\end{figure}

For comparison we also show the diagrams in the two nonclassical St{\"o}rmer
cases, see Fig.~\ref{fig:Sstable}. In the classical St{\"o}rmer
problem with $\se=\sg=0$ the derivative
$\partial_r^2 \UE |_e$ is always negative, hence there are no
stable equatorial orbits.
In the purely gravitational case stable orbits exist between the pitchfork
curve and the tangent bifurcation corresponding to $\omega_e^-$.
The curve $\omega_e^+$ always is outside the region of existence.
In the rotational St{\"o}rmer problem the pitchfork curves and the
existence curves coincide at $\omega=1$. The stable region for $\delta < 0$
is between $\omega_e^-$ and 1, while for $\delta > 0$ it is between
1 and $\omega_e^+$.
Comparing these pictures with Fig.~\ref{fig:Estable}
one clearly observes that for large $|\delta|$ (i.e.\ small mass)
the systems behaves like the rotating case, while for small
$|\delta|$ (i.e.\ large mass) the behaviour is dominated by gravity
and looks like the gravitational case.
In the gravitational case there exist stable pro- and retrograde
orbits for any $\delta$; the system is symmetric with respect to
change of sign of $\omega$ and $\delta$.
In the rotational case this symmetry is broken and both pro- and 
retrograde orbits only exist for
negative charge. Trying to interpolate between the large and small $\delta$
behaviour the full system creates an interval 
$[\delta_e^-,\delta_e^+]$ of charge-to-mass
ratios for which no stable equatorial circular orbits exist.

\subsection{Nonequatorial Orbits}
\begin{figure}
\fig{   \centerline{\epsfig{file=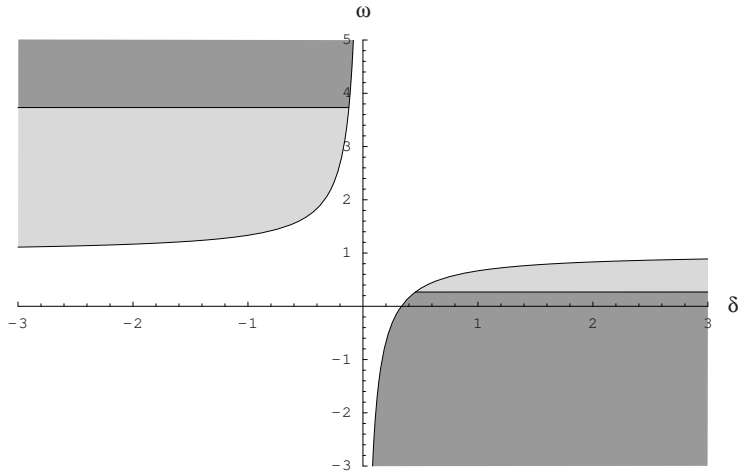,width=4in}}}
		\caption{\small{Regions of stable halo orbits are
dark. Overlay Fig.~\ref{fig:Hrcon}}}
     \label{fig:Hstable}
\end{figure}

In Ref [8] 
the stability of halo orbits was analyzed by examining the zeros of a
quintic in $\rho$.
Here we obtain similar results more directly.
For nonequatorial orbits the calculation is essentially the
same as for their equatorial cousins, except that all three
second derivatives are nonzero and we have to calculate the determinant
and the trace of the Hessian, which turn out to be
\begin{eqnarray*}
	\text{det}D^2U_{\rm eff}|_h &=& \sg \frac{2\omega^2
(\omega^2-4\omega\se+\se^2)(\sg+3\delta(\omega-\se))}
	{3\delta(\omega-1)^3}  \\
	\text{tr}D^2U_{\rm eff}|_h &=& -\sg
\frac{\omega^2(13\omega^2-16\omega\se+4\se^2)}
	{12\delta(\omega-\se)^3} -
	\frac{2(\omega-2\se)^2(\sg+3\delta(\omega-\se))}{3r_h(\omega)(\omega-\se)^2} 
\,.
\end{eqnarray*}
The determinant vanishes in three cases:
1) For $\omega=0$, which again for $\delta\not=1$ does not correspond
to finite orbits.
Since $\omega$ appears squared there is no change in stability when
orbits go through infinity.
2) In the case that the last factor is zero, which reproduces the condition
       $\omega = \omega_{PF} $.
3) There are two new critical $\omega$ given by the remaining factor as
\[
	\omega_h^\pm = \se (2 \pm \sqrt{3}) \, .
\]
These  frequency values correspond to tangent bifurcations
of halo orbits. The corresponding horizontal lines
are also shown in Fig.~\ref{fig:Estable}.
A pair of stable and unstable orbits is created
for this frequency. The lines only extend up to the
intersection with the pitch fork curve, which occurs at
\[
      \delta_h^\pm = \frac\sg\se \frac{1\pm\sqrt{3}}{6} \, .
\]
For $\delta < 0$ the upper curve
extends up to $\delta_h^-$, the lower curve is valid
for $\delta > 0$ and extends from $\delta_h^+$ to
infinity. For $\delta < 0$ halo orbits have to be above the
pitchfork line. Inserting into the invariants of the Hessian we
find that they are stable if they are above $\omega = \omega_h^+$
and unstable otherwise. In the unstable family there occurs
a maximum in radius at $\omega=2$. Otherwise the radius is a
monotonous function of $\omega$.
For $\delta > 0$  stability is reversed: orbits exist
below the pitchfork line and are stable if $\omega<\omega_h^-$.
At passage through $\omega=0$ the radius goes to infinity,
so that (sufficiently) positively charged retrograde halo orbits exist for
all large radii. Hence we obtain the following $\omega$ ranges of existence
of stable halo orbits:
\begin{eqnarray*}
	\delta < \delta_-        & : &  \omega > \omega_+ \\
\delta_- < \delta < 0     & : &  \omega > \omega_{PF}   \\
     0 < \delta < \delta_+   & : &  \omega < \omega_{PF}   \\
     \delta > \delta_+       & : &  \omega < \omega_-
\end{eqnarray*}
A simple way to characterize Fig.~\ref{fig:Hstable} is to say that halo orbits
with frequencies too close to synchronous are unstable. However, the range
of unstable frequencies is centered around orbits with $\omega=2$, i.e.
orbits that go around twice for one revolution of the planet. Halo orbits
with frequencies further away than $\sqrt{3}$ from this are stable.
Note that the equatorial orbits behave approximately in the opposite
way. For them only orbits with small $\omega$ are stable (except for small
$\delta$). This can also be interpreted in terms of the pitchfork bifurcation:
once equatorial orbits become too fast, they become unstable and create
stable nonequatorial orbits.
The corresponding picture for the gravitational St{\"o}rmer problem is not
shown, because it is trivial: in this case {\em every} halo orbit is stable,
as can be seen from the above expressions for determinant and trace of
the Hessian.

\section{Stable Halo Orbits in Space}

Considering Fig.~\ref{fig:Hrcon} we see that the curves of constant
$r$ and $\theta$ transversely
intersect each other in the regions of existence. This means that the
transformation from
$(r,\theta)$ to $(\delta,\omega)$ is invertible, which we will now show.
Instead of transforming to spherical coordinates $(r,\theta)$ we
directly transform
to cylindrical coordinates.
Equations (\ref{eq:r3is}) and (\ref{eq:s2is}) can be considered as a
transformation from $(z,\rho) = (r\cos\theta,r\sin\theta)$ to
$(\delta,\omega)$.
For each of the four types of orbits distinguished by pro/retrograde and
positive/negative charge this is a global transformation because the
Jacobian is
\[
     \det \frac{\partial(z,\rho)}{\partial(\delta,\omega)} =
\frac{2}{9\delta\omega^3r\sin\theta\cos\theta} \,,
\]
which is only singular when $\omega$, $\delta$, $r$, or $\sin2\theta$ is zero.
We already know that the latter two are only zero at the boundaries of the
valid region in $(\delta,\omega)$ space.
The inverse of the transformation is given by
\begin{eqnarray}
\label{eq:omofrt}
      \omega^2 &=& \frac23 \frac1{r^3\sin^2\theta} \\
\label{eq:delofrt}
      \delta   &=& \frac13 \frac1{(\se-\omega)\sin^2\theta} \, .
\end{eqnarray}
The first is a generalization of  Kepler's 3rd law for halo orbits,
which surprisingly
is independent of the electric field. Compared to the usual law it has
effective frequency  $\sqrt{3/2}\omega\sin\theta$.
   From the second equation the corresponding
charge to mass ratio can be calculated.
These two equations give a precise prediction about what dust particles
with what velocities should be observable at a given nonequatorial
position $(r,\theta)$.

\begin{figure}
\fig{   \centerline{ \epsfig{file=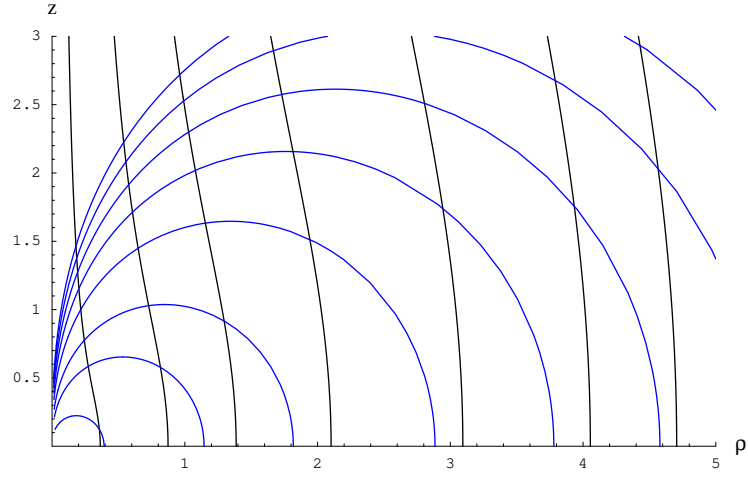,width=4in} }}
\caption{\small{$(\delta,\omega)$ grid in $(\rho,z)$ space for 
halo orbits in the gravitational St{\"o}rmer problem.
(top left to right $\pm\omega = -2-\sqrt{3},-1,-0.5,-2+\sqrt{3},-0.15,-0.1,-0.08$,
bottom $\pm\delta = 0.1,0.5,1,2,3,4,5,6$)}}
\label{fig:HSrhoz}
\end{figure}

\begin{figure}
\fig{   \centerline{\epsfig{file=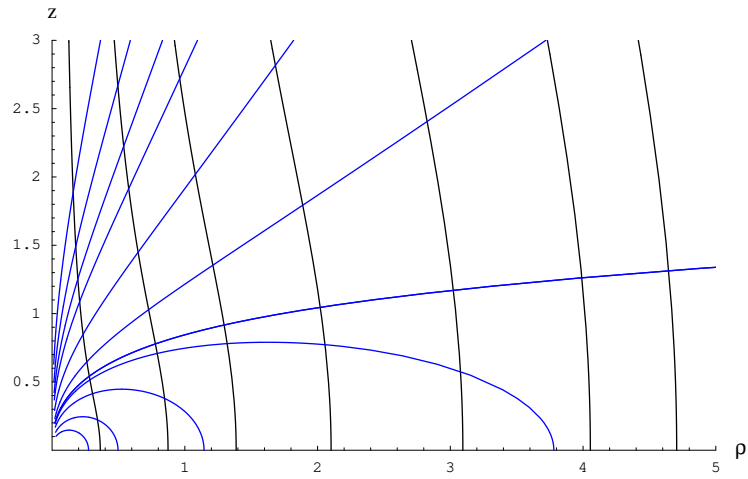,width=4in}}}
\caption{\small{$(\delta,\omega)$ grid in $(\rho,z)$ space for
    retrograde positive halo orbits. ($\omega$ as in Fig.~\ref{fig:HSrhoz},
bottom left to right, then up
$\delta = 0.05,0.1,0.2,0.3,\frac{1}{3}, 0.5,1,2,3,5,10$)}}
\label{fig:Hrhozrp}
\end{figure}

\begin{figure}
\fig{   \centerline{\epsfig{file=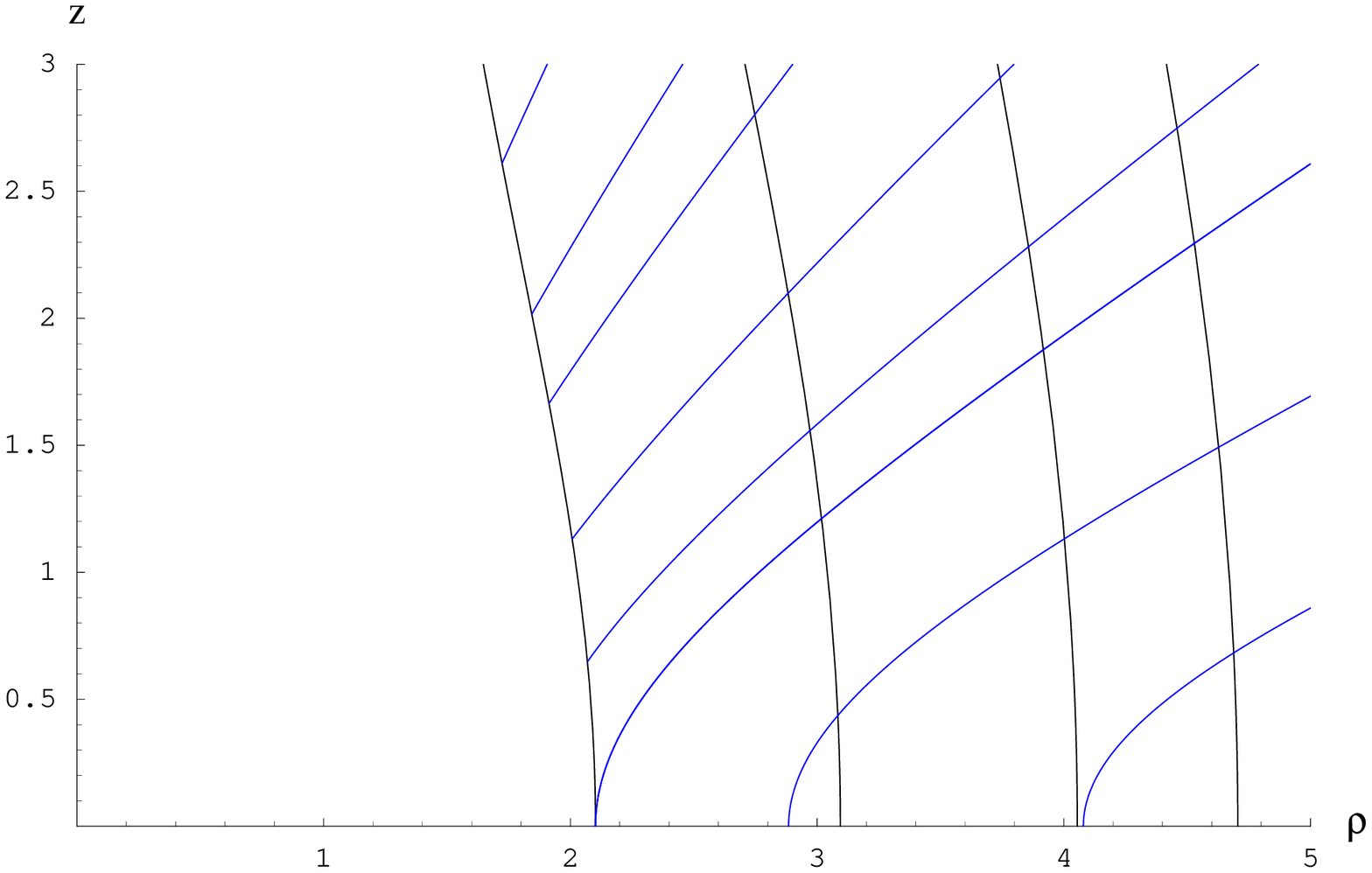,width=3in} $\quad$
  \epsfig{file=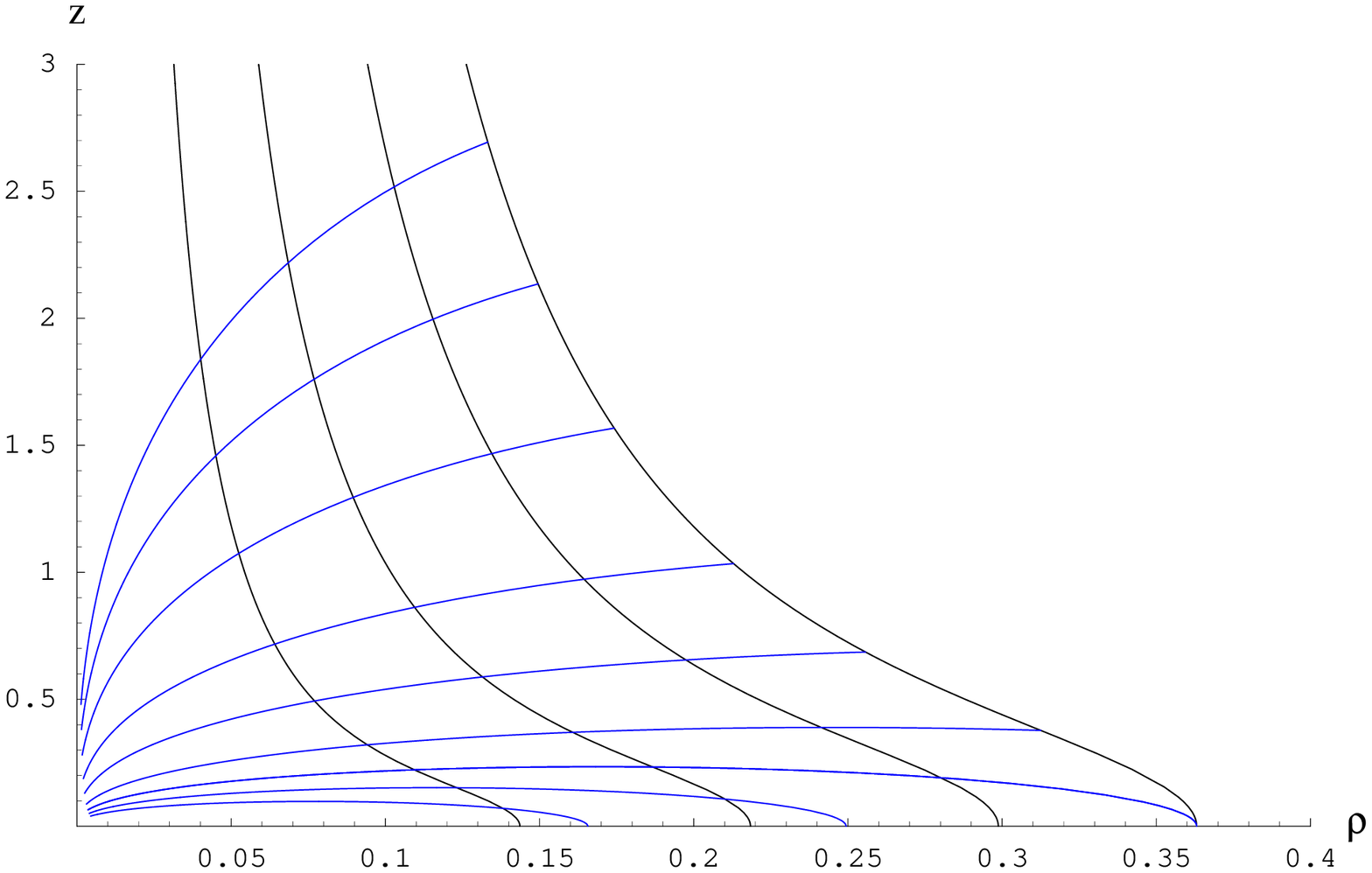,width=3in}}}
\caption{\small{$(\delta,\omega)$ grid in $(\rho,z)$ space for
a) prograde positive ($\omega = 2-\sqrt{3},0.15,0.1,0.08$,
        left top to bottom $\delta = 1.5, 1, 0.8, 0.6,
0.5,(1+\sqrt{3})/6, 0.4, 0.37$), and
b) prograde negative ($\omega = 15,8,5,2+\sqrt{3}$,
       right top to bottom $\delta = -50, -25, -10, -3, -1, -0.3,
(1-\sqrt{3})/6, -0.06, -0.03$) halo orbits.}}
\label{fig:Hrhoz}
\end{figure}

To get an idea about what particles to expect at what position we now
plot the curves of constant $\omega$ and $\delta$ on the $\rho$-$z$ plane.
The simplest way to do this is to use (\ref{eq:r3is},\ref{eq:s2is}) to generate
a parametric form of these curves in the $\rho$-$z$ plane:
\[
      (\rho,z) = (r_h\sin\theta_h, r_h\cos\theta_h).
\]
In the gravitational St{\"o}rmer case the formulas are
\[
     (\rho,z) = \left( -\frac{2\delta}{\omega} \right)^{1/3} \left(
\sqrt{-1/(3\delta\omega)}, \sqrt{1+1/(3\delta\omega)} \right).
\]
$\delta$ and $\omega$ have to be restricted to the range of
existence, respectively stability,
which is the same in the present case. The resulting diagram is shown in
Fig.~\ref{fig:HSrhoz}.
Because there is no coupling to the rotation, prograde and retrograde
orbits are
the same up to the sign of $\omega$.

For the full system there is an additional region of prograde orbits
with $\delta > 0$,
see Fig.~\ref{fig:Hrhoz}. In this case only
$(\delta,\omega)$ values from the stable
regions are taken to draw the grids. This is the reason for the cutoffs in the
prograde case. Note that all the retrograde halo orbits are stable,
and therefore
for every point in the $(\rho,z)$ plane there exists a unique halo orbit.

Note that all four figures share the same set of lines $\omega={\rm
const}$. This
is a result of the fact that the generalized 3rd Kepler's law (\ref{eq:omofrt})
is independent of $\se$, $\delta$, and independent of the sign of $\omega$.
Using the transformation from $(r,\theta)$ to
$(\rho,z)$ we can convert this equation into
\[
      z^2 = \frac94 \left( \frac\omega\rho \right)^4 - \rho^2 \,,
\]
which is the explicit form of the curves $\omega={\rm const}$ in all
four figures.
A similar explicit form for the curves $\delta = {\rm const}$ can only be
obtained with $\se = 0$. Then we find $z^2 =
(-\sqrt{6}\delta\rho)^{4/5} - \rho^2$.
With $\se=1$ the curves in $(\rho,z)$ are described by a polynomial
of degree 5 in $z^2$ and $\rho^2$, so that the above parametric form is
the most convenient representation.

Our most important conclusion is that the dependence on $\omega$ is
the same with or without the electric field. The distribution of
grain sizes as given by the curves of constant $\delta$ is
significantly changed. The changes are fairly small for stable
retrograde positive orbits. In both cases they exist at any point in
space. The prograde negative orbits need quite high angular velocity
and only survive close to the $z$ axis. Stable prograde positive halo
orbits do not exist at all without an electric field. With the field
they need to have a minimal distance of a little more then twice
the synchronous radius, and $\delta$ must be around $(1+\sqrt{3})/6$
in order to be able to be close to the planet.
It follows that retrograde positive orbits are the most likely candidates
for halo orbits. 

\section{Discussion}
We have calculated explicit equilibrium and stability conditions for 
arbitrary circular orbits in
an axisymmetric combination of gravitational, magnetic and 
corotational electric fields.
The equilibrium and stability boundaries were conveniently parametrized by the
charge-to-mass ratio
$\delta$ 
and the orbital frequency $\omega$.
The individual effects of planetary gravitational field, magnetic field
and corotational electric field on the existence and stability of charged
dust grain orbits were elucidated.

Our principal result is that halo orbits cannot exist without inclusion of
gravitational forces. Without the corotational electric field
all halo orbits are stable.
The distribution of orbital frequencies of stable halo orbits in
space is the same with- and without the corotational electric field, which is
the content of a generalized Kepler's 3rd law (\ref{eq:omofrt}).
The inclusion of the corotational electric field alone does not
give halo orbits at all. Adding it to the gravitational field
does not have a strong effect on positive retrograde orbits, which are
still all stable. It destabilizes negative prograde orbits
with small frequencies.
Adding the corotational electric field has a surprisingly strong 
effect on the character of
both equatorial and nonequatorial (halo) orbits.  In particular, 
prograde positively charged
halos require a corotational electric field for their very existence.

For halo orbits lying several Saturn radii above the equatorial plane  the typical surface potential of a dust grain is expected
to be around $+5 V$, due to the low plasma density there and resultant dominant photoelectric charging. If stable
retrograde grains are present, even the very small grains predicted by our theory should be detected by the CDA experiment on board the
Cassini orbiter due to arrive in 2004.

\end{document}